\documentclass{article}

\def\xxinput#1{\input#1}

\xxinput{vsolj02.sty}

\usepackage[dvipdfmx]{graphicx}
\usepackage[comma,colon]{natbib}

\usepackage[OT2,T1]{fontenc}

\def\cite{\citealt}
\setcitestyle{aysep={}}

\newcounter{author}
\def\altaffilmark#1{$^{#1}$}
\def\altaffiltext#1{$^{#1}$\,}

\def\authorcount#1#2{{\refstepcounter{author}\label{#1}
                     \altaffiltext{\ref{#1}}{#2}}}

\begin{document}

\begin{center}

\title{Eclipse observations of V838 Her (Nova Her 1991) during nova eruption}

\author{
        Taichi~Kato\altaffilmark{\ref{affil:Kyoto}}
}

\authorcount{affil:Kyoto}{
     Department of Astronomy, Kyoto University, Sakyo-ku,
     Kyoto 606-8502, Japan \\
     \textit{tkato@kusastro.kyoto-u.ac.jp}
}

\end{center}

\begin{abstract}
\xxinput{abst.inc}
\end{abstract}

   V838 Her (Nova Her 1991) was discovered by Matsuo Sugano
on 1991 March 24.781 (JD 2448340.281) at a photovisual
magnitude of 5.4 and by George Alcock on 1991 March 25.19
(JD 2448340.69) at a visual magnitude of
5 \citep{sug91v838heriauc5222}.
Although this was a naked-eye nova at its peak,
very few people saw the nova when it was visible to
the naked eyes or with small binoculars.
Due to the rainy or cloudy weather in Japan following
the Sugano's discovery, the first visual observation reported
to the VSOLJ database\footnote{
   $<$http://vsolj.cetus-net.org/database.html$>$.
} from Japan was 10.2~mag on 1991 April 1
by Hiroaki Narumi.  Infrared photometry and spectroscopy, and
optical spectroscopy confirming a very rapidly expanding nova
were reported in \citet{har91v838heriauc5223}.
At that time, this nova showed Fe\textsc{ii} emission lines.
\citet{wag91v838heriauc5227} obtained optical spectra and
also detected He\textsc{i}, N\textsc{ii} and other lines.
\citet{wag91v838heriauc5227} noted that the spectrum was
strikingly similar to that of Nova V1500 Cyg (fastest classical
nova known at that time and later shown as a nova eruption
from a polar) obtained on 1975 September 6--7
and suggested that the nova eruption in V838 Her occurred
on a strongly magnetized white dwarf.

   V1500 Cyg is a famous naked-eye nova which erupted in
1975.  Upon knowing this nova in a television news program,
I immediately got out to find a bright nova changing the shape
of the constellation.  The object was even suspected to be
a supernova from a large amplitude of more than 18~mag
(see e.g., \cite{lin75v1500cyg}).  No nova of comparable apparent
brightness has been recorded since V1500 Cyg.
P. Tempesti reported periodic
variations with a period of 3.2~hr already when the nova
declined to $V$=6 ($\sim$4~mag below the optical peak;
report in \cite{koz75v1500cygiauc2834})
and confirmed it again at $V$=6.5
(report in \cite{gie75v1500cygiauc2841}).
R. Koch and C. Ambruster also detected the same short-period
variations (reports in \cite{deveg75v1500cygiauc2837,
har75v1500cygiauc2839}).  \citet{sem75v1500cygibvs1058}
determined the period to be 0.1410~d (=3.38~hr).
The period was then observed to decrease to a mean
value of 0.1384~d \citep{tem76v1500cygibvs1098,
sem76v1500cygibvs1157,you77v1500cyg}.
\citet{chi77v1500cyg} remarked a resemblance to TT Ari
and considered the hot spot illuminating an overlying
shell or shells.  \citet{pat78v1500cyg} reported
an increase of this period in 1977.  Although
\citet{pat78v1500cyg} stated that the period and phase
variations have a common origin in the central object,
which powers the expanding nebula, the origin of
the variations remained unknown [see also \citet{kle79v1500cyg}].
Although \citet{pat79v1500cyg} [and later by \citet{kal87v1500cyg}]
reported the refined period, the mechanism of the variations
was not yet clarified.
In the meantime, \citet{cam76v1500cyg} recorded profile
variations on a time scale of hours of the H$\alpha$ emission
during the early decline phase and suggested a model considering
the light-travel-time effects.  \citet{hut77v1500cygspec}
proposed a searchlight illumination-type model and was
refined further in \citet{hut78v1500cygspec}.
\citet{hut79v1500cygspec} pointed out
the similarity of the light amplitude, phasing,
velocity amplitude and phasing and the phasing of emission-line
strengths to the polar AM Her, although intrinsic linear
polarization had not yet been detected
(\cite{kem76v1500cyg,mcl76v1500cygpolari,kem77v1500cyg};
these observations were apparently aimed to detect
a signature of non-spherical expansion and were
apparently not related to the possibility of a polar).
\citet{sto88v1500cyg} finally detected circular polarization
and optical cyclotron emission, confirming the strongly
magnetized nature of the white dwarf.  \citet{sto88v1500cyg}
suspected that coupling between the white dwarf and
the expanded envelope and the interaction of the orbiting
secondary star were responsible for the asynchronism and
photometric variations.

\begin{figure*}
\begin{center}
\includegraphics[width=16cm]{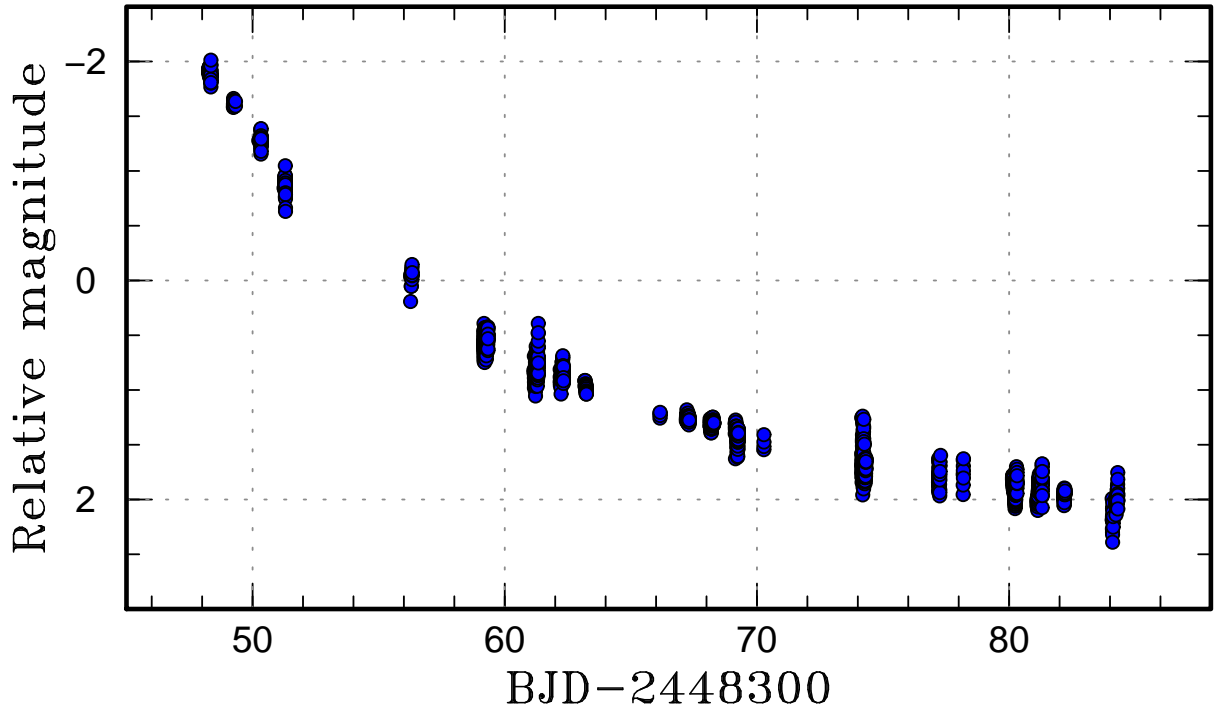}
\caption{
   Light curve of V838 Her in 1991 April--May.
   Magnitudes are given relative to the comparison star.
}
\label{fig:lc}
\end{center}
\end{figure*}

   This story of V1500 Cyg from 1975 to 1988 was still fresh
in my mind in 1991, and the suggestion of a nova on
a strongly magnetized white dwarf by
\citet{wag91v838heriauc5227} strongly attracted my attention.
I was a master course student of astronomy then, and immediately
wrote a proposal to use the 60-cm telescope at Ouda Station,
Kyoto University \citep{Ouda}.\footnote{
   This ``historical'' telescope was translocated to
   the Museum of Astronomical Telescopes was established in Sanuki,
   Kagawa, Japan ($<$https://www.telescope-museum.com/telescope/$>$).
}
The time-allocating committee of Ouda Station was, however,
skeptical whether I could obtain any scientifically meaningful
result by time-resolved photometry of a nova in eruption.  This
was understandable --- the V1500 Cyg case required more than
10 years to understand even by top experts of thie field,
and this showed how unpredictable to detect periodic variations
from a nova in eruption.
I was, however, allocated some telescope time sharing with
another student.  As people who have visited or lived in Japan
in this season (April--May) would know well, the weather
is dominated by frequently passing fronts, and rain.
It is usually very difficult to encounter ``photometric''
nights in this season, particularly for a visiting astronomer
staying for observation only for a short time.
These fronts and rain, however, bring migratory birds from
the south, and the epoch of my observations incidentally
matched the best season for meeting arriving summer migrants.
Even if the night was ruined by clouds, the chorus of birds
at dawn was not what I had heard in urban life, and
this experience was sufficient to bring me fascination
to the world of birds (e.g., \cite{kat22stageA}).
Even without detailed knowledge about birdsong,
I could immediately recognize the very characteristic song of
Japanese Paradise Flycatcher (\textit{Terpsiphone atrocaudata})\footnote{
   The reason why I was already familiar with this bird was
   that it had been introduced as an ``astronomy fan among birds''.
   Its Japanese name is ``sank\={o}ch\={o}'', which can be
   literally translated to ``three-light bird''.
   The three lights here are the Moon (``tsuki'' in Japanese),
   the Sun (``hi'' in one of expressions in Japanese) and
   stars (``hoshi'').  The bird sings ``tski-hi-hoshi hoi-hoi-hoi''
   \citep{bra09BirdsofEastAsia}.  I had learned this literal
   expression of the birdsong in a popular astronomy magazine.
   This species is a summer migrant to Japan and the exceptionally
   long tails of male birds not only attract female birds
   (the long tail is therefore a target of sexual selection) but also
   birdwatchers.  This species once became rare in the 1990s
   (around my observations at Ouda Station),
   together with other summer migrants.
   Deforestation in the wintering grounds and big Asian Forest Fires
   in 1997--1998 ($<$http://datazone.birdlife.org/sowb/casestudy/in-indonesia-human-initiated-fires-are-responsible-for-massive-losses-of-rainforest-$>$)
   were suspected to be some of the causes.
   The population has apparently increased
($<$https://www.yamashina.or.jp/hp/yomimono/hanshokubunpu\_chosa\_report.html$>$, $<$https://www.bird-atlas.jp/news/j\_bird\_atlas\_2016-21.pdf$>$)
   and we can now sometimes hear the voice in the suburbs of Kyoto.
}
--- what a surprise!

   Let's move to astronomical observations.  Instead of detecting
V1500 Cyg-like variations, I unexpectedly discovered eclipses
(report in \cite{dop91v838heriauc5262}).
These observations were probably one of the first time-resolved
CCD photometry of novae in eruption \citep{VSNET}.\footnote{
   See also the statement by \citet{sch22usco} in relation to
   the eruption of U Sco in 2010.
}
The data were obtained using a CCD camera
(Thomson TH~7882, 576 $\times$ 384 pixels, on-chip
2 $\times$ 2 binning adopted) attached to the Cassegrain focus
of the 60-cm Ritchey-Chr{\'e}tien telescope
at Ouda Station, Kyoto University.  I mainly used $V$ band
for time-resolved photometry.  I also obtained some $I$-band
data and initially narrow and wide band interference filters
(if I correctly remember, they were for H$\beta$ and Ca lines
to study the line strengths) used by another student with
whom I shared some nights.  The frames were analyzed
using a custom C code running on NEC PC9801 personal computers.
The details of the observations, however, were lost or
relocated to a place difficult to find after more than
30 years.  The only relatively easily accessible data
were $V$-band measurements posted to the VSOLJ database
and differential magnitudes stored in my computer.
At that time, the images were stored in
magneto-optical disks, which were expensive for us
to keep all the images.  Some images should be still present,
but I'm not aware where the media are placed and are
these media should be difficult to access even if they
are still present.  For this reason, I only use $V$-band
measurements posted to the VSOLJ database
and differential magnitudes stored in my computer.
The data in the VSOLJ database were converted to
real (not differential) magnitudes, and the difference
from the differential magnitudes indicates that I used
GSC 1034.3147 (Gaia DR3 4504548475874019840, Gaia $BP$=12.46
and $RP$=10.98: \cite{GaiaDR3}) as the comparison star with
a magnitude of $V$=11.80, as judged from the field of
view of the CCD.
I performed aperture photometry, and a star close to
V838 Her (Gaia DR3 4504548029197366272, Gaia $BP$=14.74
and $RP$=13.81: \cite{GaiaDR3}) contaminated the results.
In this paper, I use differential magnitudes.
As already stated, the weather in April--May was very variable
and there were many observations affected by clouds.
Outlier data (usually more than 0.1-mag different
from the rest) have been removed (one can refer to
the original data in the VSOLJ database).  This manual
removal was unavoidable since the data were often affected
by passages of clouds (different parts of the images were
differently affected), rather than photon noise-limited.
The data used here are summarized in table \ref{tab:log}.
The magnitudes are given relative to the comparison.

\xxinput{obs.inc}

   I show the overall $V$-band light curve in 1991 April--May
in figure \ref{fig:lc}.  Although I observed this object in later
seasons and detected eclipses, they are not included since
the nova faded below the nearby star and the spatial resolutions
of the images were rather insufficient for reliable PSF photometry.

I used the same ephemeris as in \citet{ing92v838her}
\begin{equation}
\mathrm{Min(BJD}) = 2448369.227(1) + 0.297635(6) E
\label{equ:ecl}
\end{equation}

\begin{figure*}
\begin{center}
\includegraphics[width=16cm]{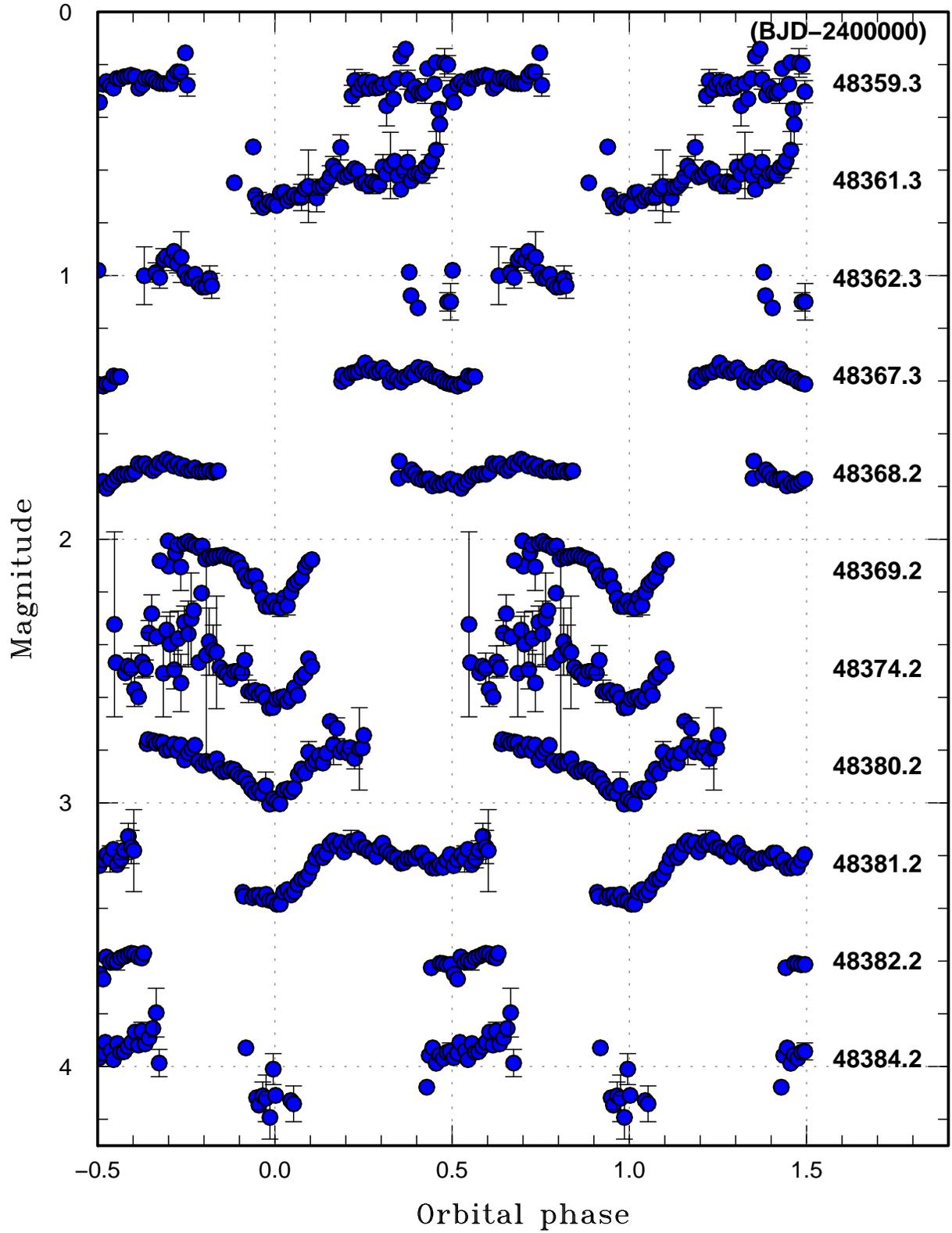}
\caption{
   Variation of the orbital profiles of V838 Her.
   The data were binned to 0.01 phase.
}
\label{fig:orb}
\end{center}
\end{figure*}

\begin{figure*}
\begin{center}
\includegraphics[width=16cm]{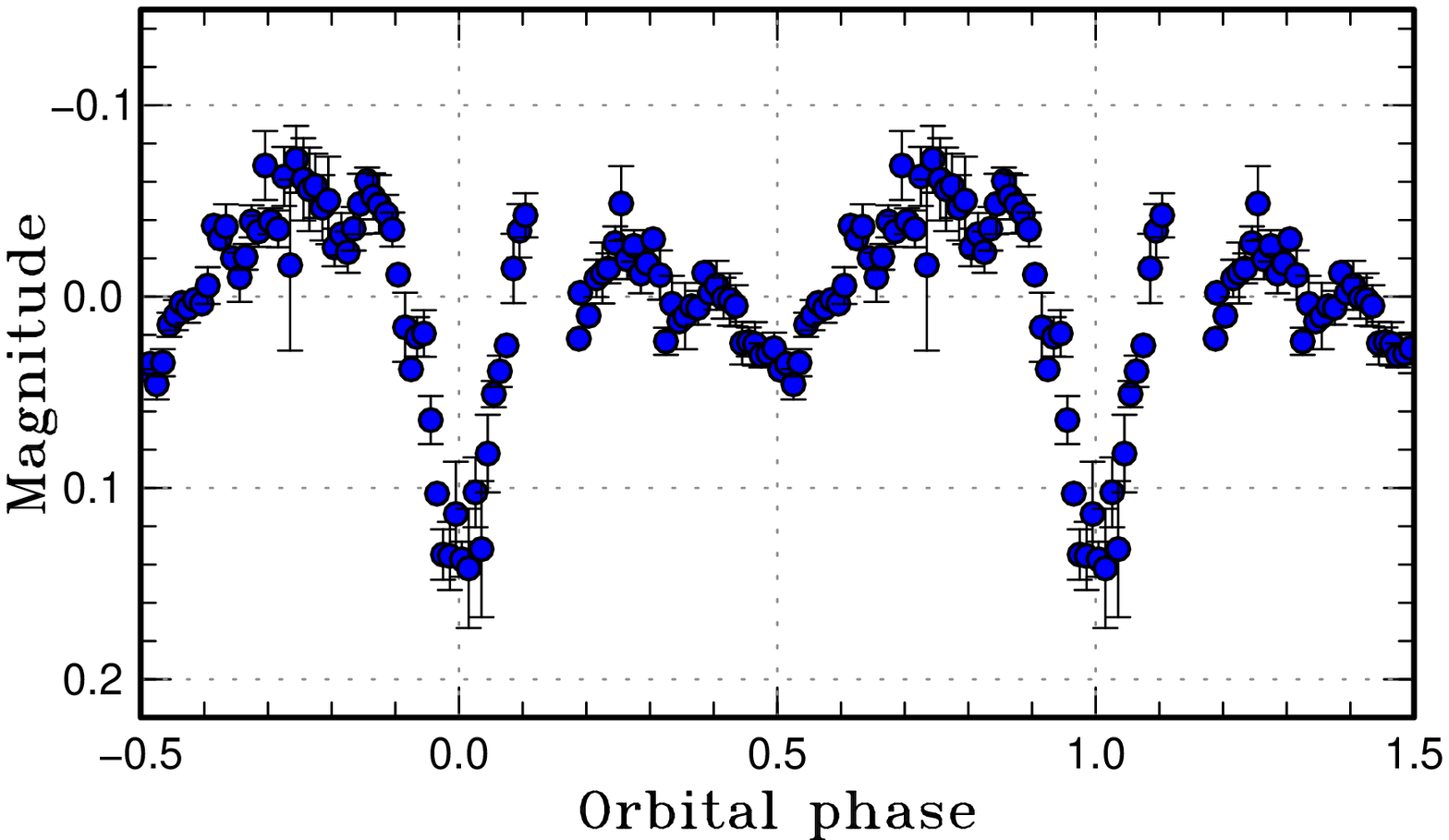}
\caption{
   Phase-averaged light curve of V838 Her on the initial three
   nights (1991 April 20--22) when eclipses appeared.
   The data were binned to 0.01 phase.
}
\label{fig:phave}
\end{center}
\end{figure*}

   The most important part of the observations showing
the appearance of eclipses is given in figure \ref{fig:orb}.
Before the initial epoch shown in this figure, there was no
hint of orbital modulations.  Although \citet{lei92v838her}
suggested the presence of an eclipse already on 1991 April 13
(BJD 2448360.59, at $+$20~d after the optical peak at 2448340.5),
our data on BJD 2448361.3 ($+$20.8~d) seems to
exclude an eclipse with a depth reported in \citet{lei92v838her}.
Considering that \citet{lei92v838her} recorded only
the expected ingress phase of an eclipse and not the egress
part, it looks premature to conclude the presence of
an eclipse before 1991 April 14 (object at $V$=12.5).
A safe conclusion is that eclipses started to appear sometime
between 1991 April 14 (BJD 2448361) and April 22 (BJD 2448369).
A phase-averaged light curve of the combined three consecutive
nights (BJD 2448367.1--2448369.3, $+$26.6--28.8~d) is shown
in figure \ref{fig:phave}.
On these nights, the object was around $V=13.1$,
8~mag below the optical peak and 7~mag above the quiescence.
As judged from this orbital profile, a secondary minimum appears
to have already present in addition to the primary eclipse.  
Eclipses were probably present on April 20 (BJD 2448367).
There was no suggestion of a secondary eclipse on April 14.

   \citet{lei92v838her} reported the presence of an accretion
disk three weeks after the nova eruption.  I here examine whether
the light curve can be expressed by an eclipse of an accretion
disk.  In modeling the eclipse light curve, I used an orbital
inclination ($i$) of 78--90$^\circ$ \citep{szk94v838herqzaur}.
The estimate of the primary (white dwarf) mass
by \citet{szk94v838herqzaur} was too small for a very fast nova,
and I used 1.35~$M_\odot$ from \citet{kat09v838her} instead.
The secondary mass was assumed to be 0.86~$M_\odot$
considering that the secondary is an unevolved main-sequence
star and was extrapolated from the table in \citet{kni06CVsecondary,
kni07CVsecondaryerratum}.  This value is not very different
from the one (0.73--0.75~$M_\odot$) in \citet{szk94v838herqzaur}.
The duration of the observed eclipse (0.2 orbital phase)
could be reproduced for $i$=78--90$^\circ$
assuming a flat, standard accretion disk (yes, this is
an awfully rough approximation) with a size limited by
tidal truncation.  The eclipse depth was 3.6~mag for
$i$=90$^\circ$ and 2.0~mag for $i$=78$^\circ$.
The observed depth (figure \ref{fig:phave}) of 0.14~mag.
roughly suggests the fractional contribution from the disk
to be 0.04 ($i$=90$^\circ$) to 0.07 ($i$=78$^\circ$).
Considering that the object was $V=13.1$ around these
observations, the apparent brightness of the disk can be
estimated to be $V$=16.6 ($i$=90$^\circ$) and
$V$=16.0 ($i$=78$^\circ$).  Using the distance modulus of
12.2(4)~mag and $E(B-V)$=0.53(5) by \citet{kat09v838her},
converted to $A_V$=1.6(2),
the brightness of the disk corresponds to
$M_V$=$+$2.8(4) ($i$=90$^\circ$) and $+$2.2(4) ($i$=78$^\circ$).
These values are $\sim$2~mag brighter than novalike systems
observed by Gaia \citep{abr20CVGaia}.  The difference becomes
even larger if the geometrical effect of the high orbital inclination
of this system is taken into account.
These observed values, however, would be explained by
a strongly irradiated disk.
I used a geometrically thin disk for modeling, which is probably
not a very good approximation of the real disk, if present,
and these values and discussions would better be regarded
as a rough approximation.

   The secondary appears on day $+$9 according in the model
by \citet{kat09v838her}.  The presence of eclipses 18~d after
the appearance of the secondary appears to be consistent.
\citet{lei92v838her} gave an upper limit of $\sim$0.05~mag
for the secondary eclipse.  The present data, however, suggest
a detectable secondary eclipse with a duration comparable
to the primary eclipse (figure \ref{fig:phave}).
The depth was $\sim$0.03~mag.

   Let's compare with modern observations of the eclipsing,
fast recurrent nova U Sco.  During its 2010 eruption,
\citet{wor10uscoiauc9114} suggested that an accretion disk
had already been re-established 7~d after the optical peak
based on appearance of flickering. 
U Sco was around $V$=12.5 then, which was $\sim$100 times
brighter than the object in quiescence.  The observed
amplitude of 0.2~mag was too large ($\sim$20 times of
the quiescent luminosity) to be considered as genuine
flickering, which is produced by the variable release of
the gravitational potential at the hot spot.
\citet{mun10uscoibvs5930} casted doubts on the interpretation
by \citet{wor10uscoiauc9114} due the absence of such variations
at 15.7~d after the optical peak.  \citet{sch11uscoecl}
interpreted such variations in the early phase of decline
as flares (of unknown nature).
\citet{sch11uscoecl} suggested from eclipse
mapping that the light source at the primary was spherically
symmetric 15--26~d after the optical peak and
that it was a rim-bright disk 26--41~d after
the peak.  Although re-establishment of the accretion disk
7~d after the optical peak would have been too early
in U Sco, I can make a comparison whether the disk could
be re-established in V838~Her at $+$27~d.
\citet{dra10usco} considered the case of U Sco and
a mass-transfer rate of 10$^{-8}$ to 10$^{-6}$ $M_\odot$ yr$^{-1}$
was sufficient to build up a disk in 7~d, if the dynamic pressure
of the radiatively-driven outflow from the evolving central object
could be overcome.  One should also note that the pre-eruption
surface densities of the disk (and, consequently, the disk mass)
are not absolutely required when the disk is re-established.
Before the eruption, the disk should have been
in a hot, ionized state due to viscous heating, which requires
high surface densities.  After the nova eruption when
the white dwarf is sufficiently hot, the disk can be
maintained in hot, ionized state due to radiation even if
the surface density is low and a lower-mass disk is expected
to explain the observation.
The pre-eruption mass-transfer rate of V838~Her
is expected to be comparable to that of a novalike star:
10$^{-8}$ $M_\odot$ yr$^{-1}$ or larger \citep{kni11CVdonor}.
Irradiation by the primary would probably increase the mass-transfer
rate and the condition given by \citet{dra10usco} appears
to have easily been achieved, if only the dynamic pressure
of the radiatively-driven outflow from the evolving central object
could be overcome.  I therefore could not find a strong argument
against the presence of an accretion disk in V838~Her at $+$27~d.

   The presence of black body emission at $\sim$20000~K
was reported in infrared observations of U Sco \citep{eva23uscoIR},
This component was attributed to the irradiated secondary
or a combination with free-free emissions from the nebula.
In the case of V838 Her, the secondary filling the Roche lobe
and having a temperature of 20000~K corresponds to
$M_V$=$+$4.7 (compare with $M_V$=$+$7.1 for a non-irradiated
secondary: \cite{kat09v838her}).  The secondary with this
temperature comprises 0.7\% of the luminosity of the nova
at $V$=13.1.  Although this contribution (0.007~mag) is
much lower than the depth (0.03~mag) of the secondary minimum
I recorded, the contribution of the secondary
may become an observable one assuming an even higher
temperature (0.02~mag for a full eclipse of a 40000~K object).
Considering the observational uncertainty, the secondary eclipse
suggested by my observations could have been a real feature
arising from the eclipse of the secondary.
The fastest known classical nova, but non-eclipsing,
V1674 Her showed orbital variations starting
from $+$5~d, 4~mag below the optical peak, often accompanied
by secondary minima \citep{pat22v1674her}, in which a transient
luminous donor (=irradiated secondary, as in I my suggestion
for V838 Her) was given as a hypothesis.

\section*{Supplementary Data}

   The $V$-band differential data used in this analysis
is given as a plain text v.bjd.

\section*{Acknowledgements}

This work was supported by JSPS KAKENHI Grant Number 21K03616.
I am grateful to Dr. Ryuko Hirata, who was my supervisor
when I was a student and assisted me when writing a report
to IAU Circ. No. 5262, and to Thomas Djamaluddin, who shared
observation and lived together at Ouda Station when
I was a beginner observer.

\section*{List of objects in this paper}
\xxinput{objlist.inc}

\section*{References}

We provide two forms of the references section (for ADS
and as published) so that the references can be easily
incorporated into ADS.

\newcommand{\noop}[1]{}\newcommand{\hyphalt}{-}

\renewcommand\refname{\textbf{References (for ADS)}}

\xxinput{v838heraph.bbl}

\renewcommand\refname{\textbf{References (as published)}}

\xxinput{v838her.bbl.vsolj}


\begin{thebibliography}{}

\bibitem[{Abril} et~al.(2020)]{abr20CVGaia}
  {Abril}, J., {Schmidtobreick}, L., {Ederoclite}, A., \& {L{\'o}pez-Sanjuan},
  C.\ 2020, MNRAS, 492, L40 (arXiv:1912.01531)

\bibitem[{Brazil}(2009)]{bra09BirdsofEastAsia}
  {Brazil}, M.\ 2009, {Birds of East Asia}
 (Princeton: Princeton University Press)

\bibitem[{Campbell}(1976)]{cam76v1500cyg}
  {Campbell}, B.\ 1976, ApJ, 207, L41 (https://doi.org/10.1086/182174)

\bibitem[{Chia} et~al.(1977)]{chi77v1500cyg}
  {Chia}, T.~T., {Milone}, E.~F., \& {Robb}, R.\ 1977, Ap\&SS, 48, 3
  (https://doi.org/10.1007/BF00643035)

\bibitem[{de Vegt} et~al.(1975)]{deveg75v1500cygiauc2837}
  {de Vegt}, C., {et~al.}\ 1975, IAU Circ., 2837, 2

\bibitem[{Dopita} et~al.(1991)]{dop91v838heriauc5262}
  {Dopita}, M., {Ryder}, S., {Vassiliadis}, E., {Kato}, T., \& {Hirata}, R.\
  1991, IAU Circ., 5262, 1

\bibitem[{Drake} and {Orlando}(2010)]{dra10usco}
  {Drake}, J.~J., \& {Orlando}, S.\ 2010, ApJ, 720, L195 (arXiv:1007.2810)

\bibitem[{Evans} et~al.(2023)]{eva23uscoIR}
  {Evans}, A., {Banerjee}, D.~P.~K., {Woodward}, C.~E., {R.}, T., {Geballe},
  {Gehrz}, R.~D., {Page}, K.~L., \& {Starrfield}, S.\ 2023, MNRAS, in press
  (arXiv/2304.13508)

\bibitem[{Gaia Collaboration} et~al.(2022)]{GaiaDR3}
  {Gaia Collaboration}, {et~al.}\ 2022, A\&A, (arXiv:2208.00211)

\bibitem[{Gieren} et~al.(1975)]{gie75v1500cygiauc2841}
  {Gieren}, W., {et~al.}\ 1975, IAU Circ., 2842, 1

\bibitem[{Harevich} et~al.(1975)]{har75v1500cygiauc2839}
  {Harevich}, V., {et~al.}\ 1975, IAU Circ., 2839, 1

\bibitem[{Harrison} et~al.(1991)]{har91v838heriauc5223}
  {Harrison}, T.~E., {et~al.}\ 1991, IAU Circ., 5223, 1

\bibitem[{Hutchings}(1979)]{hut79v1500cygspec}
  {Hutchings}, J.~B.\ 1979, ApJ, 230, 162 (https://doi.org/10.1086/157072)

\bibitem[{Hutchings} et~al.(1978)]{hut78v1500cygspec}
  {Hutchings}, J.~B., {Bernard}, J.~E., \& {Margetish}, L.\ 1978, ApJ, 224, 899
  (https://doi.org/10.1086/156440)

\bibitem[{Hutchings} and {McCall}(1977)]{hut77v1500cygspec}
  {Hutchings}, J.~B., \& {McCall}, M.~L.\ 1977, ApJ, 217, 775
  (https://doi.org/10.1086/155624)

\bibitem[{Ingram} et~al.(1992)]{ing92v838her}
  {Ingram}, D., {Garnavich}, P., {Green}, P., \& {Szkody}, P.\ 1992, PASP, 104,
  402 (https://doi.org/10.1086/133012)

\bibitem[{Kaluzny} and {Semeniuk}(1987)]{kal87v1500cyg}
  {Kaluzny}, J., \& {Semeniuk}, I.\ 1987, Acta Astron., 37, 349

\bibitem[{Kato} et~al.(2009)]{kat09v838her}
  {Kato}, M., {Hachisu}, I., \& {Cassatella}, A.\ 2009, ApJ, 704, 1676
  (arXiv:0909.1506)

\bibitem[{Kato}(2022)]{kat22stageA}
  {Kato}, T.\ 2022, VSOLJ Variable Star Bull., 89, (arXiv:2201.02945)

\bibitem[{Kato} et~al.(2004)]{VSNET}
  {Kato}, T., {Uemura}, M., {Ishioka}, R., {Nogami}, D., {Kunjaya}, C., {Baba},
  H., \& {Yamaoka}, H.\ 2004, PASJ, 56, S1 (arXiv:astro-ph/0310209)

\bibitem[{Kemp} and {Rudy}(1976)]{kem76v1500cyg}
  {Kemp}, J.~C., \& {Rudy}, R.~J.\ 1976, ApJ, 203, L131
  (https://doi.org/10.1086/182036)

\bibitem[{Kemp} et~al.(1977)]{kem77v1500cyg}
  {Kemp}, J.~C., {Sykes}, M.~V., \& {Rudy}, R.~J.\ 1977, ApJ, 211, L71
  (https://doi.org/10.1086/182344)

\bibitem[{Kleine} and {Kohoutek}(1979)]{kle79v1500cyg}
  {Kleine}, T., \& {Kohoutek}, L.\ 1979, A\&A, 76, 133

\bibitem[{Knigge}(2006)]{kni06CVsecondary}
  {Knigge}, C.\ 2006, MNRAS, 373, 484 (arXiv:astro-ph/0609671)

\bibitem[{Knigge}(2007)]{kni07CVsecondaryerratum}
  {Knigge}, C.\ 2007, MNRAS, 382, 1982
  (https://doi.org/10.1111/j.1365-2966.2007.12206.x)

\bibitem[{Knigge} et~al.(2011)]{kni11CVdonor}
  {Knigge}, C., {Baraffe}, I., \& {Patterson}, J.\ 2011, ApJS, 194, 28
  (arXiv:1102.2440)

\bibitem[{Kozai} et~al.(1975)]{koz75v1500cygiauc2834}
  {Kozai}, Y., {et~al.}\ 1975, IAU Circ., 2834, 1

\bibitem[{Leibowitz} et~al.(1992)]{lei92v838her}
  {Leibowitz}, E.~M., {Mendelson}, H., {Mashal}, E., {Prialnik}, D., \&
  {Seitter}, W.~C.\ 1992, ApJ, 385, L49 (https://doi.org/10.1086/186275)

\bibitem[{Lindegren} and {Lindgren}(1975)]{lin75v1500cyg}
  {Lindegren}, L., \& {Lindgren}, H.\ 1975, Nature, 258, 501
  (https://doi.org/10.1038/258501a0)

\bibitem[{McLean}(1976)]{mcl76v1500cygpolari}
  {McLean}, I.~S.\ 1976, MNRAS, 176, 73
  (https://doi.org/10.1093/mnras/176.1.73)

\bibitem[{Munari} et~al.(2010)]{mun10uscoibvs5930}
  {Munari}, U., {Dallaporta}, S., \& {Castellani}, F.\ 2010, IBVS, 5930, 1
  (arXiv:1003.2870)

\bibitem[{Ohtani} et~al.(1992)]{Ouda}
  {Ohtani}, H., {et~al.}\ 1992, Mem.\ Faculty of Sciences,\ Kyoto Univ.,\
  Series A of Physics, Astrophysics, Geophysics and Chemistry, 38, 167

\bibitem[{Patterson}(1978)]{pat78v1500cyg}
  {Patterson}, J.\ 1978, ApJ, 225, 954 (https://doi.org/10.1086/156562)

\bibitem[{Patterson}(1979)]{pat79v1500cyg}
  {Patterson}, J.\ 1979, ApJ, 231, 789 (https://doi.org/10.1086/157244)

\bibitem[{Patterson} et~al.(2022)]{pat22v1674her}
  {Patterson}, J., {et~al.}\ 2022, ApJ, 940, L56 (arXiv:2207.00181)

\bibitem[{Schaefer}(2022)]{sch22usco}
  {Schaefer}, B.~E.\ 2022, MNRAS, 516, 4497 (arXiv:2206.14231)

\bibitem[{Schaefer} et~al.(2011)]{sch11uscoecl}
  {Schaefer}, B.~E., {et~al.}\ 2011, ApJ, 742, 113 (arXiv:1108.1214)

\bibitem[{Semeniuk}(1975)]{sem75v1500cygibvs1058}
  {Semeniuk}, I.\ 1975, IBVS, 1058, 1

\bibitem[{Semeniuk} et~al.(1976)]{sem76v1500cygibvs1157}
  {Semeniuk}, I., {Kruszewski}, A., \& {Schwarzenberg-Czerny}, A.\ 1976, IBVS,
  1157, 1

\bibitem[{Stockman} et~al.(1988)]{sto88v1500cyg}
  {Stockman}, H.~S., {Schmidt}, G.~D., \& {Lamb}, D.~Q.\ 1988, ApJ, 332, 282
  (https://doi.org/10.1086/166652)

\bibitem[{Sugano} et~al.(1991)]{sug91v838heriauc5222}
  {Sugano}, M., {Kosai}, H., {Alcock}, G.~E.~D., {Hurst}, G.~M., {McNaught},
  R.~H., {Harrison}, T., \& {Buczynski}, D.\ 1991, IAU Circ., 5222, 1

\bibitem[{Szkody} and {Ingram}(1994)]{szk94v838herqzaur}
  {Szkody}, P., \& {Ingram}, D.\ 1994, ApJ, 420, 830
  (https://doi.org/10.1086/173607)

\bibitem[{Tempesti}(1976)]{tem76v1500cygibvs1098}
  {Tempesti}, P.\ 1976, IBVS, 1098, 1

\bibitem[{Wagner} et~al.(1991)]{wag91v838heriauc5227}
  {Wagner}, R.~M., {Bertram}, R., {Ali}, B., \& {Starrfield}, S.~G.\ 1991, IAU
  Circ., 5227, 2

\bibitem[{Worters} et~al.(2010)]{wor10uscoiauc9114}
  {Worters}, H.~L., {Eyres}, S.~P.~S., {Rushton}, M.~T., \& {Schaefer}, B.\
  2010, IAU Circ., 9114, 1

\bibitem[{Young} et~al.(1977)]{you77v1500cyg}
  {Young}, P.~J., {Robinson}, E.~L., {Africano}, J., {Ferland}, G.~J., \&
  {Woodman}, J.\ 1977, PASP, 89, 37 (https://doi.org/10.1086/130067)

\end{thebibliography}
\end{document}